\documentclass[12pt,a4paper,twoside]{article}
\usepackage[dvips]{graphicx}
\usepackage[dvips]{graphicx}
\usepackage{cite}
\newcommand{\epem}   {\ensuremath{\mathrm{e^+e^-}}}

\newcommand{\PZ}     {\ensuremath{\mathrm{Z}}}

\newcommand{\dedx}   {\ensuremath{\mathrm{d}E/\mathrm{d}x}}
\newcommand{\nqb}    {\ensuremath{\langle n_q\rangle}}
\newcommand{\nub}    {\ensuremath{\langle n_{\mathrm{u}}\rangle}}
\newcommand{\ndb}    {\ensuremath{\langle n_{\mathrm{d}}\rangle}}
\newcommand{\nsb}    {\ensuremath{\langle n_{\mathrm{s}}\rangle}}
\newcommand{\ncb}    {\ensuremath{\langle n_{\mathrm{c}}\rangle}}
\newcommand{\nbb}    {\ensuremath{\langle n_{\mathrm{b}}\rangle}}
\newcommand{\Ks}     {\ensuremath{\mathrm{K^0_S}}}
\newcommand{\Kpm}    {\ensuremath{\mathrm{K^{\pm}}}}
\newcommand{\hx}     {\ensuremath{\mathrm{high}\,x_E}}
\newcommand{\ssbar}  {\ensuremath{\mathrm{s\bar s}}}
\newcommand{\apm}[2]{\ensuremath{^{\textstyle #1}_{\textstyle #2}}}
\renewcommand{\Huge}{\huge}
\begin{document}
%
%
\begin{titlepage}
\begin{center}{\large   EUROPEAN ORGANIZATION FOR NUCLEAR RESEARCH
}\end{center}\bigskip
\begin{flushright}
       CERN-EP-2000-128   \\ 
       October 04, 2000
\end{flushright}
\bigskip\bigskip\bigskip\bigskip\bigskip
\begin{center}
              {\Huge\bf Charged Multiplicities in Z Decays into u, d, and s Quarks}
\end{center}
\bigskip\bigskip
\begin{center}
{\LARGE The OPAL Collaboration}
\end{center}
\bigskip\bigskip
\bigskip
\begin{center}{\large  Abstract}\end{center}
%
About 4.4 million hadronic decays of Z bosons, recorded by the OPAL detector at
LEP at a centre-of-mass energy of around $\sqrt{s}= 91.2$\,GeV, are used to
determine the mean charged particle multiplicities for the three light quark
flavours.  Events from primary u, d, and s quarks are tagged by selecting 
characteristic particles which carry a large fraction of the beam energy. The 
charged particle multiplicities are measured in the hemispheres opposite to 
these particles. An unfolding procedure is applied to obtain these
multiplicities for each primary light quark flavour. This yields
\begin{displaymath}
\nub=\mathrm{17.77\pm 0.51\,^{\displaystyle +0.86}_{\displaystyle -1.20}}\,,\;
\ndb=\mathrm{21.44\pm 0.63\,^{\displaystyle +1.46}_{\displaystyle -1.17}}\,,\;
\nsb=\mathrm{20.02\pm 0.13\,^{\displaystyle +0.39}_{\displaystyle -0.37}}\,,\;
\end{displaymath}
where statistical and systematic errors are given.  The results for 
\nub\ and \ndb\ are almost fully statistically anti-correlated. 
Within the errors the result is consistent with the 
flavour independence of the strong interaction for the particle
multiplicities in events from the light up, down, and strange quarks.
%
%
\bigskip\bigskip\bigskip\bigskip
\bigskip\bigskip
\begin{center}{\large
(Submitted to European Physical Journal C)
}\end{center}
\end{titlepage}
%
%
\begin{center}{\Large        The OPAL Collaboration
}\end{center}\bigskip
\begin{center}{
G.\thinspace Abbiendi$^{  2}$,
C.\thinspace Ainsley$^{  5}$,
P.F.\thinspace {\AA}kesson$^{  3}$,
G.\thinspace Alexander$^{ 22}$,
J.\thinspace Allison$^{ 16}$,
G.\thinspace Anagnostou$^{  1}$,
K.J.\thinspace Anderson$^{  9}$,
S.\thinspace Arcelli$^{ 17}$,
S.\thinspace Asai$^{ 23}$,
D.\thinspace Axen$^{ 27}$
% ...
}\end{center}\bigskip
\bigskip
$^{  1}$School of Physics and Astronomy, University of Birmingham,
Birmingham B15 2TT, UK
\newline
$^{  2}$Dipartimento di Fisica dell' Universit\`a di Bologna and INFN,
I-40126 Bologna, Italy
\newline
$^{  3}$Physikalisches Institut, Universit\"at Bonn,
D-53115 Bonn, Germany
\newline
% ...
$^{  5}$Cavendish Laboratory, Cambridge CB3 0HE, UK
\newline
% ...
$^{  9}$Enrico Fermi Institute and Department of Physics,
University of Chicago, Chicago IL 60637, USA
\newline
% ...
$^{ 16}$Department of Physics, Schuster Laboratory, The University,
Manchester M13 9PL, UK
\newline
$^{ 17}$Department of Physics, University of Maryland,
College Park, MD 20742, USA
\newline
% ...
$^{ 22}$Department of Physics and Astronomy, Tel Aviv University,
Tel Aviv 69978, Israel
\newline
$^{ 23}$International Centre for Elementary Particle Physics and
Department of Physics, University of Tokyo, Tokyo 113-0033, and
Kobe University, Kobe 657-8501, Japan
\newline
% ...
$^{ 27}$University of British Columbia, Department of Physics,
Vancouver BC V6T 1Z1, Canada
\newline
% ...
Tel Aviv 69978, Israel.
\bigskip
%
%
%
\section{Introduction}
\label{sec-introduction}
The flavour independence of the strong coupling is a fundamental property of
                   quantum chromodynamics 
(QCD). A breaking of the flavour symmetry should only occur due to 
calculable mass effects. These mass effects have been observed for 
bottom quarks~\cite{bib-quarkmass-jetrates,bib-OPAL-ZPC60-397} using 
              event shapes and
jet rates in the final state of electron-positron 
              annihilation
into 
$\mathrm{b\overline{b}}$.  Another observable which can be employed 
to test 
flavour independence is the multiplicity of charged hadrons 
in jets originating from quarks of a specific flavour. 
              According to the local parton hadron duality hypothesis 
              (LPHD)~\cite{bib-LPHD}, the particle multiplicity is related
              to the gluon multiplicity inside a jet which depends on the 
              value of the strong coupling constant~\cite{bib-nlla-webber}.
A dependence of 
the multiplicity of charged hadrons on the quark flavour has been found 
for heavy quarks~\cite{bib-quarkmass-multiplicity}.  
Up to now only 
              few measurements exist on 
the flavour independence of the strong interaction 
              in the light quark sector (up, down, strange quarks)~\cite{bib-OPAL-ZPC60-397}. 
Within 
the large uncertainties due to the limited statistics, 
              flavour independence of the strong interaction is supported.

This paper presents a new, high statistics investigation of the flavour
dependence of the strong interaction. It is based on the mean charged 
multiplicity determined separately for events of primary up, down, and 
strange quarks in \epem\ annihilation at centre-of-mass energies close 
to the mass of the Z boson. 
              To identify the flavour of the primary quark in the \PZ\ decay, the
leading particle effect is exploited~\cite{bib-leading-particle-effect}. This
assumption of a correlation between the flavour of the primary quark and the
type of the hadron carrying the largest momentum has recently received further
support by measurements of the 
              SLD~\cite{bib-SLD-PRL78-3442} and 
              OPAL collaborations~\cite{bib-OPAL-ZPC76-387}.
              \par

The study presented in this paper uses three different selections 
              of events with leading \Ks\ and \Kpm\ mesons, and highly 
              energetic stable charged particles, denoted ``\hx'' in the
              following. 
              Due to the different 
              fractions
of primary up, down, and strange 
              quarks in these samples a statistical decomposition of the 
              contributions 
              from
            each of the three quark flavours is possible. 
              The 
              leading \Ks\ and \Kpm\ meson
            selections, even though 
              dominantly stemming from primary strange quarks,
            yield the separation of up and down 
              quark events since a leading \Ks\ (\Kpm) meson is rarely formed from
              a primary up (down) quark while both \Ks\ and \Kpm\ are equally likely to
              be produced in a strange quark event.
              Together with the \hx\ selection,
            containing up, down, and strange 
              quark events
            in approximately 
            equal proportions, 
              properties like the mean charged particle multiplicity in events of each 
              of these three light quark flavours can be determined by statistical 
              unfolding.
              The mean charged multiplicity is determined from all long-lived charged
              particles in the hemisphere opposite to the leading particle where the
              two hemispheres of an event are defined by the plane perpendicular to 
              the thrust axis which contains the main interaction point. Choosing
              the opposite hemisphere minimizes the bias of the measured multiplicity
              due to the high energy of the leading particle.

Before the three selections 
              of leading particles
are presented in Section~\ref{sec-selection}, the
OPAL detector and the data samples used for this study are briefly introduced
in Section~\ref{sec-detector}. The determination of the charged 
              particle
multiplicities
is discussed in Section~\ref{sec-multiplicities}. The investigation of various
sources of systematic uncertainty is detailed in Section~\ref{sec-systematics}.
              Section~\ref{sec-fitted-fractions} presents a cross-check of the 
              analysis using flavour fractions obtained from the data. The results 
              of this analysis are presented in Section~\ref{sec-result-syst}.
%
%
%
\section{The OPAL detector, data, and Monte Carlo simulation}
\label{sec-detector}
The OPAL detector has been described in detail elsewhere~\cite{bib-OPALdet}. The
analysis presented here relies mainly on the reconstruction of charged particles
in the large volume jet chamber whose performance has been presented 
              in~\cite{bib-CJperformance}. 
A solenoidal coil surrounds all tracking
detectors. It provides a field of 0.435\,T along the beam
axis\footnote{The coordinate system of OPAL has the $z$ axis along the electron
beam direction, the $y$ axis points upwards and $x$ towards the centre of the
LEP ring. The polar angle $\theta$ is measured with respect to the $z$ axis.
}. Tracks of charged particles are reconstructed with up to 159 space points. 
Their momenta in the plane transverse to the beam-axis, $p_{xy}$, can be 
determined to a precision of 
   $\sigma_{p_{xy}}/p_{xy}=\sqrt{0.02^2+(p_{xy}\cdot 0.0015/{\mathrm{GeV}} )^2}$.
This resolution degrades towards the acceptance boundary of 
              $|\cos\theta\,|\approx 0.98$. 
From the specific energy loss \dedx, which is measured from up to 159 samples with 
a resolution of $\sigma(\dedx)/(\dedx)\approx 0.035$, the type of the charged particle 
can be identified over a wide momentum range.

The analysis is based on data recorded with the OPAL detector between 1990 and
1995 comprising about 4.4 million hadronic events at centre-of-mass energies
              $\sqrt{s}$
around 91.2\,GeV (LEP~I). The events considered for this study were preselected
by a standard selection for high multiplicity
events~\cite{bib-TKMH} which relies on a minimum number of measured
              tracks of charged particles in the tracking detectors and clusters 
              of energy deposited in the electromagnetic calorimeter. 
The remaining background of 
              two-photon
processes and
$\tau$-pair events are estimated to be 
              0.07\,\%
and 0.11\,\%, respectively.

Tracks to be used for the reconstruction or identification of the leading
particles and for the determination of the charged multiplicity 
              were
required to pass the 
quality selection 
              cuts as detailed in~\cite{bib-quality}.
              These tracks had 
to have at least 20 hits in the jet 
              chamber and
a closest approach to the interaction point in the plane perpendicular 
to the beam axis of 
              $|d_0|<5$\,cm.
The track momenta 
              had to
be between 0.1 and 65\,GeV. 
              Events with $\tau$-pairs
are further suppressed by requiring at least
seven 
tracks.

              The Monte Carlo simulation comprised about 8.5 million hadronic 
              events 
              simulated
by version 7.4 of the JETSET
program~\cite{bib-JETSET} which has been tuned to describe the OPAL
data~\cite{bib-OPALtune}. The generated events were passed through a detailed
simulation of the OPAL detector~\cite{bib-GOPAL} and processed using the same
reconstruction and selection algorithm as the measured data.
%
%
%
\section{Selection of flavour enriched data samples}
\label{sec-selection}
              Every event is divided into two hemispheres by the plane perpendicular to the thrust 
              axis containing the interaction point. In each hemisphere we searched for \Ks\ and 
              \Kpm\ mesons, and highly energetic stable charged particles, whose energy 
              fraction $x_E=2E/\sqrt{s}$ 
              is above a certain threshold.
Due to the cuts, in particular on $x_E$, these three
samples are enriched in events from 
              up, down, and strange quarks in different proportions
              (see Table~\ref{tab-fractions}). The $x_E$ selection cuts were
chosen to have 
              samples of comparable size.
\subsection{$\mathbf{K^0_S}$ selection}
\Ks\ mesons were selected via their decay into two charged 
              pions using the procedure described in~\cite{bib-K0s}.
It was adapted for large momenta 
              of the \Ks\ mesons 
by dropping the specific 
              $d_0$ cut 
imposed on the two pions in the
default procedure. Any two oppositely charged particles were combined.
              The invariant mass $m_{\pi\pi}$ was calculated
              assuming the pion mass for the particles.
\Ks\ candidates were required to have
$|m_{\Ks}-m_{\pi\pi}|<60$\,MeV
              using the mass of a \Ks\ meson, $m_{\Ks}$, given in~\cite{bib-PDG}.
Photon conversions were rejected by demanding
$m_{\mathrm{ee}}>100$\,MeV if the electron mass is assigned to both tracks. 
              The impact of detector effects close to the acceptance 
              boundary is reduced by restricting the polar angle of 
              \Ks\ candidates to $|\cos\theta\,|<0.9$. 
To
enrich the sample in primary strange and down quarks and to suppress up, charm,
and bottom events, a requirement on the scaled energy of 
              the \Ks\ candidate of
$x_E>0.4$ was applied.
              Candidates with $x_E>1.07$ 
were rejected. This takes into
account the 7\,\% momentum resolution for particles with the beam energy.

\begin{figure}
\vspace*{-4mm}
\includegraphics[width=\textwidth]{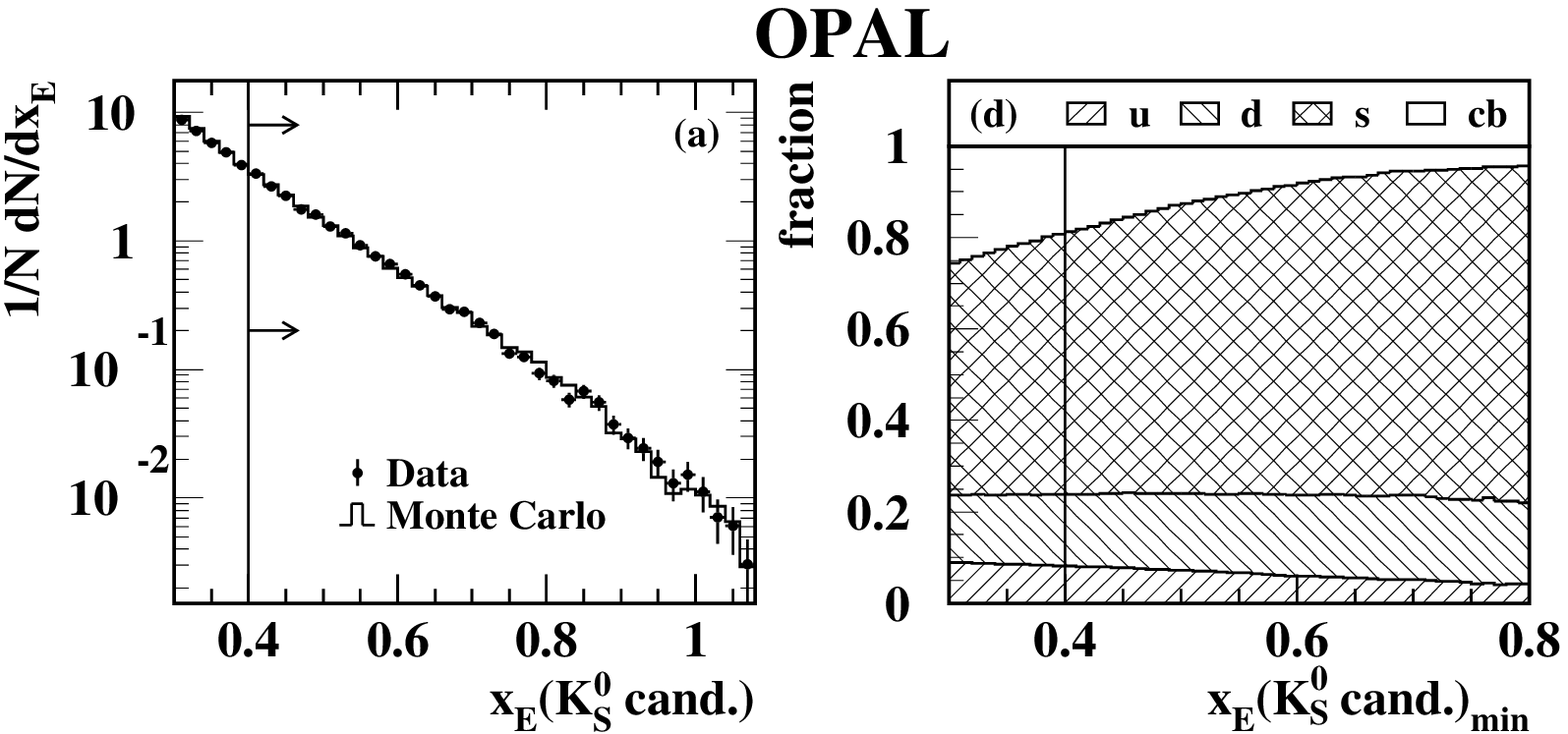}
\includegraphics[width=\textwidth]{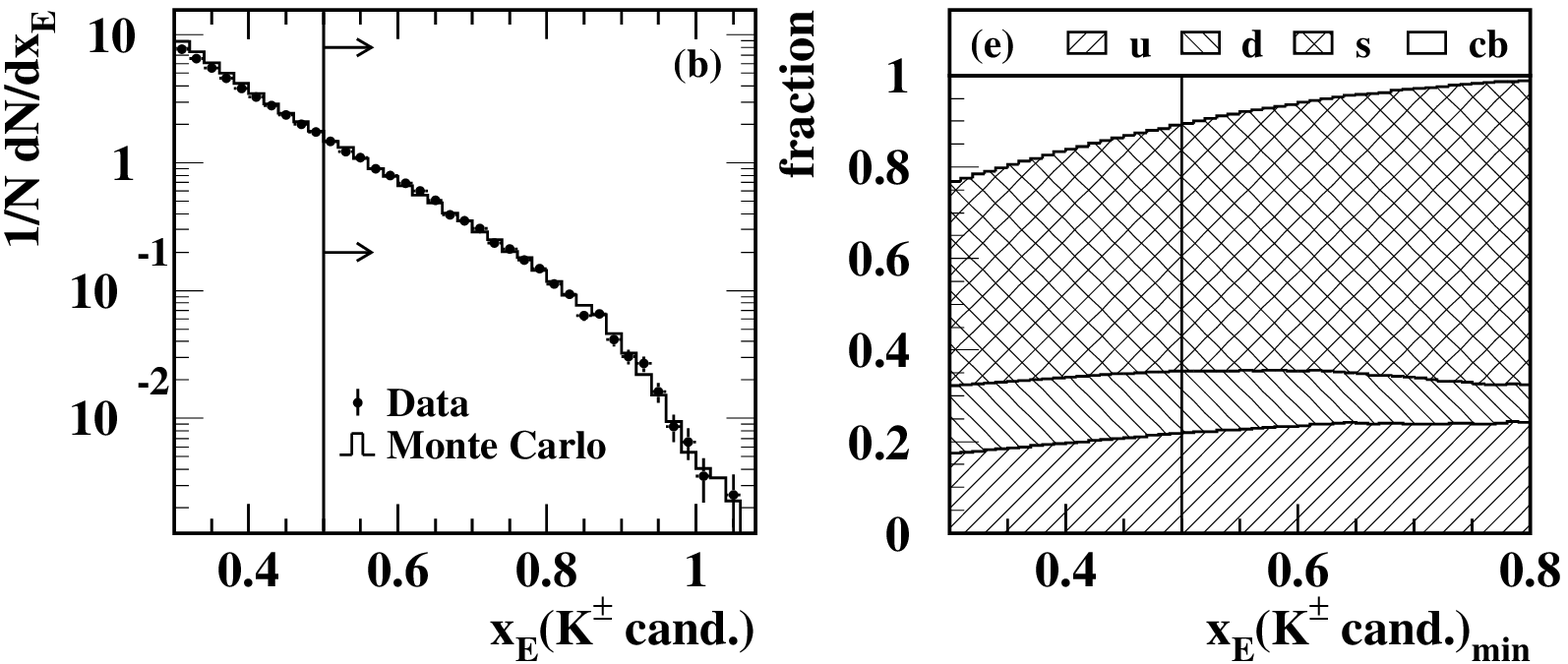}
\includegraphics[width=\textwidth]{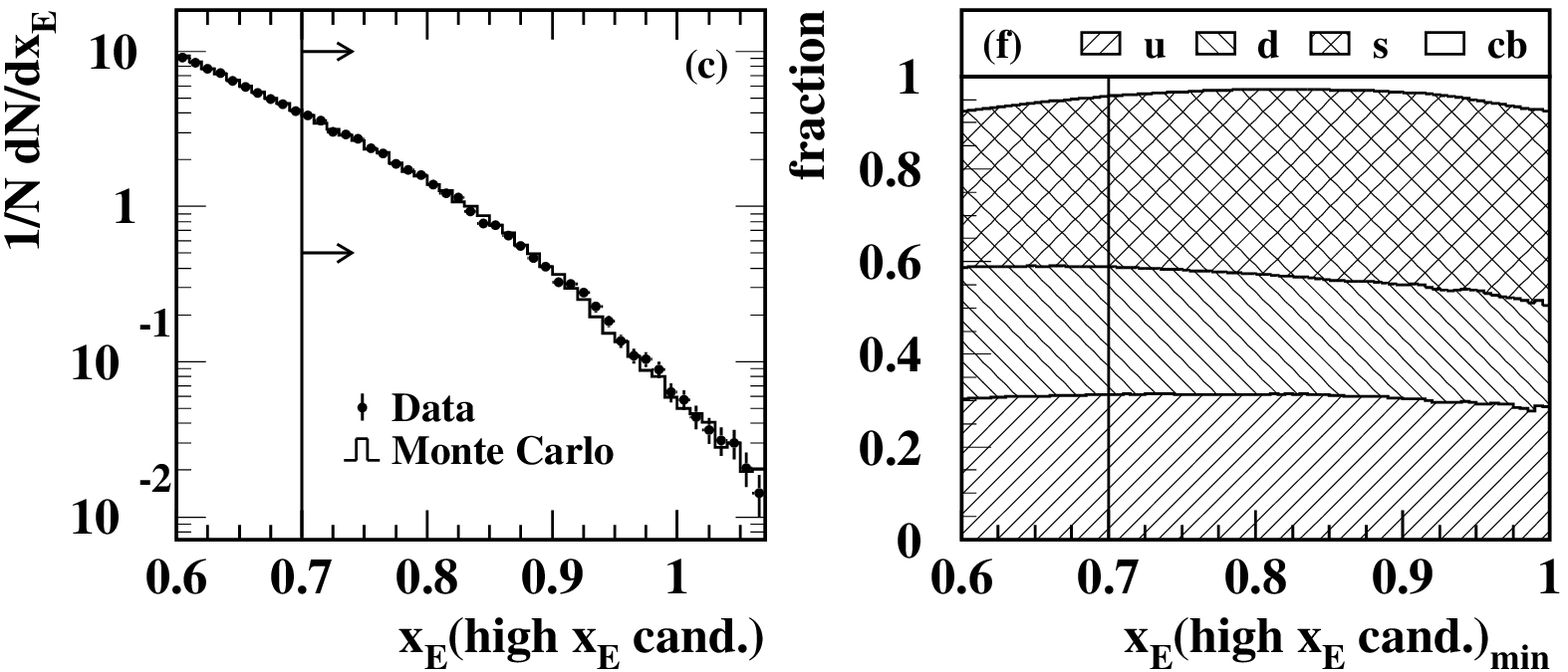}
\caption{\label{fig-spectra}
              (a)-(c)
              Observed
  energy spectra of \Ks, \Kpm, and \hx\
              candidates.
              The solid line 
              shows the Monte Carlo simulation
              normalized to the integrated luminosity of the data
              after applying the reweighting procedure described in 
              Section~\ref{sec-check-flavour-composition}.
              (d)-(f) 
              Corresponding flavour 
              fractions
  taken from the Monte Carlo 
  simulation 
  as a function of the cut on the minimum energy 
              fraction, $x_{E,\mathrm{min}}$,
  of the tag particle. The $x_E$ cuts
  chosen for the flavour enriched samples are indicated by the vertical lines.}
\end{figure}

Figure~\ref{fig-spectra} 
              (a) and (d) show
the energy spectrum 
              of the \Ks\ candidates
and the expected flavour composition of 
              the events tagged by
the \Ks\ candidates. 
Table~\ref{tab-fractions} 
              gives
the number of tagged hemispheres in the data as
well as the expected number and the flavour composition of the sample taken from
                            the
Monte Carlo simulation plus their statistical errors.
\begin{table}
\begin{center}
\begin{tabular}{|l|l||r@{$\pm$}r|r@{$\pm$}r|r@{$\pm$}r|}
\cline{2-8}
\multicolumn{1}{c|}{} & 
                        tag         & 
                                      \multicolumn{2}{|c|}{\Ks}  & \multicolumn{2}{|c|}{\Kpm} & \multicolumn{2}{|c|}{\hx}  \\
\cline{2-8}
\multicolumn{1}{c|}{} & 
                        $x_E$ range   
                                    & \multicolumn{2}{|c|}
                                                          {0.40-1.07}
                                                                 & \multicolumn{2}{|c|}
                                                                                       {0.50-1.07}
                                                                                              & \multicolumn{2}{|c|}
                                                                                                                    {0.70-1.07}
                                                                                                                            \\
\cline{2-8}
\hline
Data                   & 
                         $N^{\mathrm{tag}}$
                                           \hfill =                    & 19359 & 139 & 18979 & 138 & 27909 & 167 \\
MC expectation         & 
                         $N^{\mathrm{tag}}_{\mathrm{MC}}$
                                                         \hfill =      & 19303 &  99 & 18947 &  99 & 27845 & 119 \\
\hline
\hline
u       fraction [$10^{-2}$] & $f_{\mathrm{u}}^{\mathrm{tag}}$\hfill = &   8.3 & 0.1 &  21.7 & 0.2 &  31.4 & 0.2 \\
d       fraction [$10^{-2}$] & $f_{\mathrm{d}}^{\mathrm{tag}}$\hfill = &  15.6 & 0.2 &  13.9 & 0.2 &  27.8 & 0.2 \\
s\hfill fraction [$10^{-2}$] & $f_{\mathrm{s}}^{\mathrm{tag}}$\hfill = &  57.0 & 0.3 &  53.1 & 0.3 &  36.4 & 0.2 \\
c\hfill fraction [$10^{-2}$] & $f_{\mathrm{c}}^{\mathrm{tag}}$\hfill = &  14.6 & 0.2 &   9.0 & 0.2 &   3.0 & 0.1 \\
b       fraction [$10^{-2}$] & $f_{\mathrm{b}}^{\mathrm{tag}}$\hfill = &   4.5 & 0.1 &   2.3 & 0.1 &   1.4 & 0.1 \\
\hline
\end{tabular}
\end{center}
\caption{\label{tab-fractions}
  Number of tagged hemispheres in data and 
              in the 
              Monte Carlo simulation,
              scaled to the same integrated luminosity,
  and the flavour composition of the three samples found from the 
  simulation 
              after applying the reweighting procedure described in Section~\ref{sec-check-flavour-composition}.
  The errors are statistical only.}
\end{table}
The sizeable charm quark contribution is due to D mesons decaying into 
              \Ks.
\subsection{$\mathbf{K^{\pm}}$ selection}
Charged kaons were identified using the energy loss measurements in the jet
chamber~\cite{bib-CJperformance}. To effectively reject the contribution of
charged pions and protons, only tracks in the central region of the jet chamber,
$|\cos\theta\,|<0.72$, were considered which have at least 130 hits for
the momentum measurement and a minimum of 100 hits 
for the determination
of the energy loss. 
              Tracks
azimuthally closer than $0.5^{\circ}$ to
an anode wire plane of the jet chamber were 
              not
accepted 
              to avoid biases from
the significantly degraded resolution of both momentum and energy loss measurement 
close to these planes.

Each track 
              is assigned
five \dedx\ weights, $w_h$, 
calculated from the 
probabilities of the measured energy loss of the track to be consistent with that 
expected for a particle of type $h$, where $h$ can be 
p($\bar\mathrm{p}$), 
\Kpm, $\pi^\pm$, $\mu^\pm$ or e$^\pm$. 
              The weight is positive signed if \dedx(measured)$>$\dedx(expected) according
              to a particular hypothesis, and negative otherwise.
\Kpm\ were selected by requiring:
\begin{eqnarray*}
|w_{\mathrm{K}}| & > & \max\,(|w_{\mathrm{p}}|,|w_{\pi}|,|w_{\mu}|,|w_{\mathrm{e}}|) \\
|w_{\mathrm{K}}| & > & 0.1 \\
              \mbox{and~~~~}
|w_{\pi}|        & < & 0.1
\end{eqnarray*}
This yielded, according to the simulation, a 
composition 
              of the sample of tagged particles
of about 77\,\% kaons, 13\,\% pions, and 9.5\,\% 
              protons. The 
remainder 
              is
due to muons and charged
hyperons. It was not attempted to 
correct the \Kpm\ selection for the
pion and proton contributions
              directly. 

Figure~\ref{fig-spectra} 
              (b)
shows the energy spectrum of the \Kpm\ candidates.
To
enrich the sample in 
              events of
primary strange and up quarks, a cut on the minimum scaled
energy of $x_E>0.5$ was 
              applied as shown in Figure~\ref{fig-spectra} (e).
Badly reconstructed tracks were rejected by
requiring $x_E<1.07$. Table~\ref{tab-fractions} summarizes the number of tagged
              hemispheres, the expected number and
the primary flavour composition
taken from the 
Monte Carlo simulation.
\subsection{$\mathbf{High\,x_E}$ particle selection}
              Even though 
the fragmentation of bottom and charm quarks is harder than that of the
light quarks, the cascade decays of the b- and c-hadrons lead to many stable
particles which consequently have a significantly 
              softer
energy spectrum. Thus
the selection of 
              stable charged particles 
              carrying a high energy fraction
              $x_E$, calculated using the assumption that the
              particle is a pion,
depletes an event sample of primary charm and bottom quarks.
Besides the 
              cuts 
on the scaled energy of $0.7<x_E<1.07$, additional cuts
concerned the selection of well-reconstructed tracks which were required to have
at least 130 hits used for the reconstruction, a polar angle of
$|\cos\theta\,|<0.72$ and an azimuthal angular distance from the closest anode
wire plane of at least $0.5^{\circ}$. Moreover, the $\chi^2_{r\varphi}$ per
degree of freedom of the fit of the track to the hits in the plane perpendicular
to the beam axis had to 
              be less than 2.
This tight cut rejected
badly reconstructed tracks, including those containing kinks due to an
unrecognized decay in flight.

The energy spectrum of the \hx\ candidates and the flavour composition 
depending on the cut on the minimum energy of the \hx\ particles 
              are
shown in Figures~\ref{fig-spectra}
              (c) and (f).
The slightly rising fraction of
heavy quark events at very large $x_E$ is due to the 
              remaining
contribution of badly
reconstructed tracks in this energy region. Such tracks occur 
with the same 
              low
probability for each primary flavour. 
             \par
In Table~\ref{tab-fractions} the number of tagged hemispheres found in the data 
are given and compared with the expected number from the simulation. It should 
be pointed out that about 10\,\% of the \hx\ hemispheres are also selected with 
the \Kpm\ selection cuts. No rejection was applied, but it was checked that
eliminating the 10\,\% had no significant effect on the result of this analysis.
              The small statistical correlation was taken into account.
\subsection{Checks of the flavour composition}
\label{sec-check-flavour-composition}
A crucial ingredient in this analysis is the knowledge of the flavour
composition of the three 
              tagged event samples.
The 
              JETSET Monte Carlo simulation
              using the tuned parameters 
              of~\cite{bib-OPALtune}
predicts 
              for each of the tagged particles
in the region of high 
              values of
$x_E$ a larger differential 
cross-section $(1/N)\cdot ({\mathrm{d}}N/{\mathrm{d}}x_E)$ 
              than measured in the data.
                           This discrepancy 
              was accounted for by applying a flavour 
              independent reweighting 
              to
            each of the three 
              tagged event samples
            in the simulation. The 
              reweighting factors were 
              derived
            from the ratio of the $x_E$ spectra of the 
              tagged particles in the data and the simulation. They were parametrized by 
              a function 
              $f(x_E)=c\;\!\cdot(1-x_E)^{\delta}$ 
              whose form is motivated by the Lund symmetric fragmentation 
              function~\cite{bib-LUND-symmetric}.
              Each tag type had its own reweighting 
              function, but $\delta$ was found to be consistent for all three tags. The 
              impact of this reweighting on the flavour fractions is negligible and was 
              considered as an additional systematic uncertainty in Section~\ref{sec-systematics}.
              
              The 
              flavour composition and its sensitivity to 
              changes in the fragmentation function can be checked with measured
              data. To this end the number of events with a single and a double 
              tag have been counted and compared in measured and simulated data 
              following the proposal 
              of~\cite{bib-Letts-Maettig}. 
              Here 
              a double-tagged event has exactly 
              one tagged particle in each hemisphere,
              whereas single-tagged events have at least one tag. 
These counts depend on the sums 
              and products of the
flavour fractions.
Due to 
              the small number of tagging samples it
is not possible to find a unique solution in 
              this analysis.
However, the single-tagged and double-tagged events can be used 
to cross-check the
flavour fractions obtained from the 
Monte Carlo simulation
              using measured data only.

Double-tagged events were studied in different $x_E$ ranges of the tagged
particles. Each of the three samples was divided into two $x_E$ regions. For
each sample the lower $x_E$ region was chosen 
              so that the 
              impact from the choice of the parameters of the fragmentation function 
              is small.
The remaining part corresponds to the higher $x_E$ region 
              which is particularly sensitive to the hardness of the fragmentation 
              function.
In detail, the lower $x_E$ ranges
were chosen for the \Ks\ and \Kpm\ samples to be $0.4<x_E<0.6$, and for the \hx\
sample to be $0.6<x_E<0.7$. The higher $x_E$ regions are $0.6<x_E<1.07$ for both
kaon selections and $0.7<x_E<1.07$ for the \hx\ sample. For both the 
              lower and the higher
$x_E$ regions the number of 
              double-tagged
events were determined. 
The events
were classified according to 
              whether (a) both tags belong to the lower $x_E$ region, or (b) 
at least one tag belongs to a higher $x_E$ region.
\begin{table}[t]
\begin{center}
\begin{tabular}{|c|l||r@{$\pm$}r||r@{\hspace{0pt}}c@{\hspace{0pt}}r
                                 |r@{\hspace{0pt}}c@{\hspace{0pt}}r
                                 |r@{\hspace{0pt}}c@{\hspace{0pt}}r|}
\cline{3-13}
\multicolumn{2}{c|}{(a)} & \multicolumn{2}{|c||}
                                                {$N^{\mathrm{tag}}$} &
\multicolumn{3}{|c|}{\Ks} & \multicolumn{3}{|c|}{\Kpm} & \multicolumn{3}{|c|}{\hx} \\
\hline
                     & data          & 16236 & 127 &     23 & $\pm$ &     5 &    116 & $\pm$ &    11 &    124 & $\pm$ &    11 \\
\raisebox{1ex}[0mm][0mm]{\Ks}  
                     & MC            & 16185 &  91 &     33 & $\pm$ &     4 &    111 & $\pm$ &     8 &    123 & $\pm$ &     8 \\
\hline
                     & data          & 34952 & 187 & \multicolumn{3}{|c|}{} &    144 & $\pm$ &    12 &    340 & $\pm$ &    18 \\
\raisebox{1ex}[0mm][0mm]{\Kpm} 
                     & MC            & 35075 & 134 & \multicolumn{3}{|c|}{} &    151 & $\pm$ &    10 &    350 & $\pm$ &    14 \\
\hline
                     & data          & 49160 & 222 & \multicolumn{3}{|c|}{} & \multicolumn{3}{|c|}{} &    258 & $\pm$ &    16 \\
\raisebox{1ex}[0mm][0mm]{\hx}  
                     & MC            & 49245 & 159 & \multicolumn{3}{|c|}{} & \multicolumn{3}{|c|}{} &    273 & $\pm$ &    12 \\
\hline
\multicolumn{13}{c}{} \\
\multicolumn{13}{c}{} \\
\cline{3-13}
\multicolumn{2}{c|}{(b)} & \multicolumn{2}{|c||}
                                                {$N^{\mathrm{tag}}$} &
\multicolumn{3}{|c|}{\Ks} & \multicolumn{3}{|c|}{\Kpm} & \multicolumn{3}{|c|}{\hx} \\
\hline
                     & data          &  3123 &  56 &     14 & $\pm$ &     4 &  24 & $\pm$ &  5       &    101 & $\pm$ &    10 \\
\raisebox{1ex}[0mm][0mm]{\Ks}  
                     & MC            &  3118 &  40 &     19 & $\pm$ &     3 &  23 & $\pm$ &  3       &    113 & $\pm$ &     6 \\
\hline
                     & data          &  8099 &  90 & \multicolumn{3}{|c|}{} &      2 & $\pm$ &     1 &    197 & $\pm$ &    14 \\
\raisebox{1ex}[0mm][0mm]{\Kpm} 
                     & MC            &  7938 &  64 & \multicolumn{3}{|c|}{} &      0 &       &       &    190 & $\pm$ &     9 \\
\hline
                     & data          & 27909 & 167 & \multicolumn{3}{|c|}{} & \multicolumn{3}{|c|}{} &    419 & $\pm$ &    20 \\
\raisebox{1ex}[0mm][0mm]{\hx}  
                     & MC            & 27755 & 119 & \multicolumn{3}{|c|}{} & \multicolumn{3}{|c|}{} &    407 & $\pm$ &    12 \\
\hline
\end{tabular}
\end{center}
\caption{\label{tab-double-tags}
  Number of doubly tagged events when: (a) both tags belong to the lower $x_E$
  region, (b) at least one tag belongs to a higher $x_E$ region. The upper rows
  are the measured data. The 
              lower rows present the 
              reweighted
              simulation data (MC). Also the
              total number of tags are given for data 
              and the simulation (third column).}
\end{table}
Table~\ref{tab-double-tags} lists the double tag counts for data and simulation,
normalized to the integrated luminosity of the 
              data after the reweighting.
             Good
agreement between the double tag counts in data and simulation is 
              found for both $x_E$ regions, as
can be seen from 
              Table~\ref{tab-double-tags}. 
              This can be 
              quantified using
              the $\chi^2$ per degree of freedom of the double tag counts which is 
                                                         $3.6/8$ ($5.3/8$) 
              in the 
              higher (lower)
           $x_E$ regions.
            Since the double tag counts depend on the flavour fractions, the good 
            agreement for the absolute number of tags and for the number of double 
            tags 
              leads to the conclusion that the
              estimate of the flavour fractions by the 
              JETSET generator
              is reliable.

Different multiplicities for events from primary u, d, or s quarks 
              could imply
different fragmentation functions for these quarks. 
              To test the impact of different fragmentation 
              functions
another 
              cross-check was conducted
which considers a
different reweighting factor
              $f(x_E)=c\;\!\cdot(1-x_E)^{\delta}$
for one of the three light quarks. This
corresponds to the case of a different fragmentation function for 
              one quark flavour
with respect to the 
              other two.
The flavour dependence was 
              modelled
by allowing in turn for a different value of $\delta$ for one of
the three light quarks while maintaining as a constraint 
              in the fit of the 
              $\delta$ parameters
the agreement of both
the 
              number of
single tag counts and the $x_E$ spectra in data and simulation as before.

When this reweighting was applied to the simulation, the light flavour fractions
of the two kaon selections changed by less than 
              0.08,
while for the \hx\ selection the fractions changed by up to a factor of 
              two.
              The
comparison of the
double tag counts after the flavour dependent reweighting 
              revealed, however,
a worse
description of the data by the simulation than in the case of the flavour
independent reweighting. 
              This can be inferred
              from the $\chi^2$ per degree of freedom
              for the 
              number of double tag
            counts in the whole $x_E$ regions considered which
              increased to
                                          $27.7/7$, $13.7/7$, or $36.5/7$,
              respectively for each sample,
              when allowing for a different value of $\delta$ 
              and, hence, a different fragmentation function
            for up, down, or strange 
              quarks.
              These large values of $\chi^2$ per degree of freedom strongly disfavour 
              a flavour dependent reweighting 
              of the fragmentation functions
            and, 
              therefore,
           the substantial changes
           of the flavour fractions associated with this reweighting. This gives 
           further evidence that the flavour fractions derived from the simulation 
           can be considered to be an accurate estimate of the flavour fractions 
           in the data.
              We, therefore, use for our main result the flavour fractions obtained 
              from JETSET without application of the reweighting procedure. The
              small effect on the flavour fractions due to the reweighting will 
              yield a systematic uncertainty which is discussed in 
              Section~\ref{sec-systematics}.
%
%
%
\section{Determination of charged particle multiplicities}
\label{sec-multiplicities}
\subsection{Determination of flavour dependent multiplicities}
The flavour dependent multiplicities are defined as the average number of
charged particles with life-times $\tau>300$\,ps emerging from a decay of 
a \PZ\ into one of the three light quark flavours, u, d, or s,
             as 
             for the inclusive multiplicities 
             in~\cite{bib-OPAL-mult}.
To obtain these multiplicities from the charged particle tracks 
measured for each of the three selections, several corrections must be applied. 
In particular, the finite detector resolution and acceptance, and biases due to 
the tag selections must be accounted for.

\begin{figure}[b]
\includegraphics[width=\textwidth]{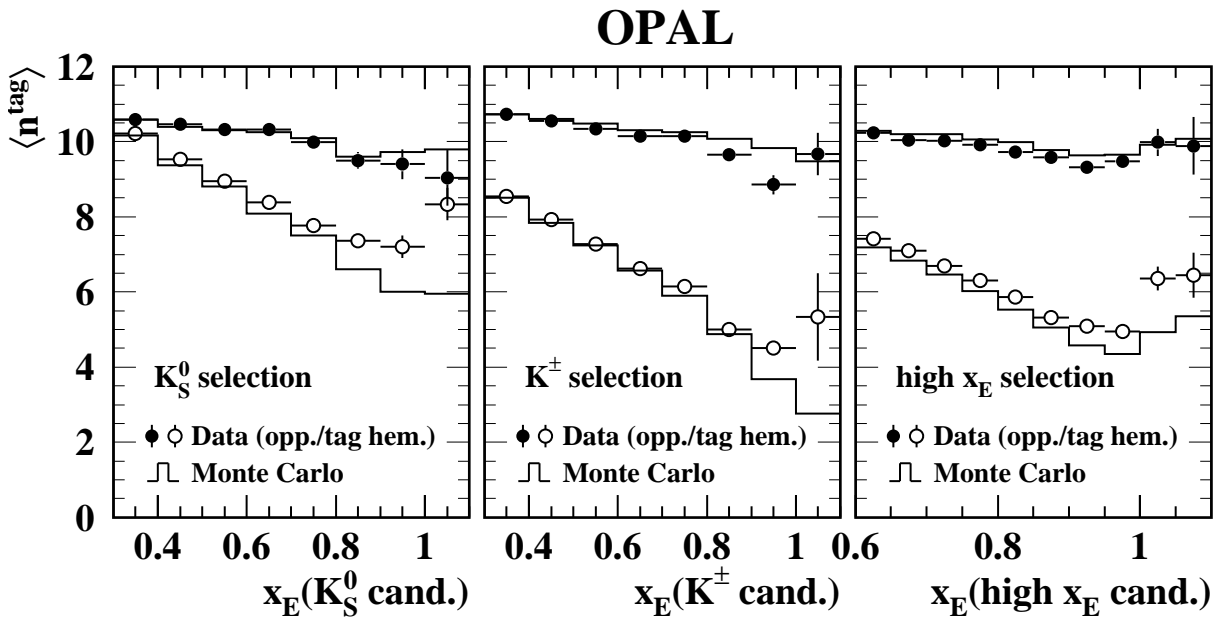}
\caption{\label{fig-hem-corr}
  Mean 
              charged multiplicity in a hemisphere
  as a function of the energy of the 
              tagged
  particles. Open circles 
  show the values obtained from the hemispheres containing the 
              tagged
  particle, full circles correspond to the opposite hemispheres.
               The histogram is the expectation from the 
               Monte Carlo simulation.
  }
\end{figure}
              The requirement of a 
              high energy of the 
              tagging 
particles in all three selections 
              affects the multiplicity in its hemisphere
              due to a reduced phase space for the production of other particles.
              In the
opposite hemisphere the effect is much smaller. This can be seen in
Figure~\ref{fig-hem-corr} where the mean 
              charged particle multiplicities per hemisphere
in the three
selections are plotted in bins of the tagging particle's energy. To keep the
bias of the multiplicity due to the 
              tagging
small we determine the multiplicities
only in the hemispheres opposite to the tagged particles. 
              \par The remaining small dependence of the multiplicity in the
              opposite hemisphere on the energy of the tagging particle 
              is due to the impact of the high energy of this particle.
              The higher the energy required of the leading particle 
              the more the phase space 
              is reduced,
               which is available for particle production,
              and, therefore, the mean charged multiplicity in the 
              opposite hemisphere.
It should also be
noted from 
              Figure~\ref{fig-hem-corr}
that the 
Monte Carlo simulation satisfactorily describes
the effects in the opposite hemisphere due to the tag, although it tends to
slightly overestimate the 
              multiplicity in particular if the energy of the tagging
              particle is very high. This is due to a slight underestimation
              of the rate of badly reconstructed tracks in the simulation 
              which particularly affects the hemisphere of the tagging 
              particle if the measured energy fraction, $x_E$, of such a track is 
              close to or above the physical limit of 1. Since the differential 
              cross-section is tiny for leading particles with $x_E$ close to 1, 
              the effect of events selected due to badly reconstructed tracks on 
              the mean charged multiplicity in the opposite hemisphere is negligible
              when considering the full range of $x_E$ used for the tagging.
              Any remaining 
              discrepancies between data and Monte Carlo simulation in
              the opposite hemisphere
are taken
as a source of systematic error, which is treated in Section~\ref{sec-systematics}.
              \par
\begin{figure}[t]
\includegraphics[width=\textwidth]{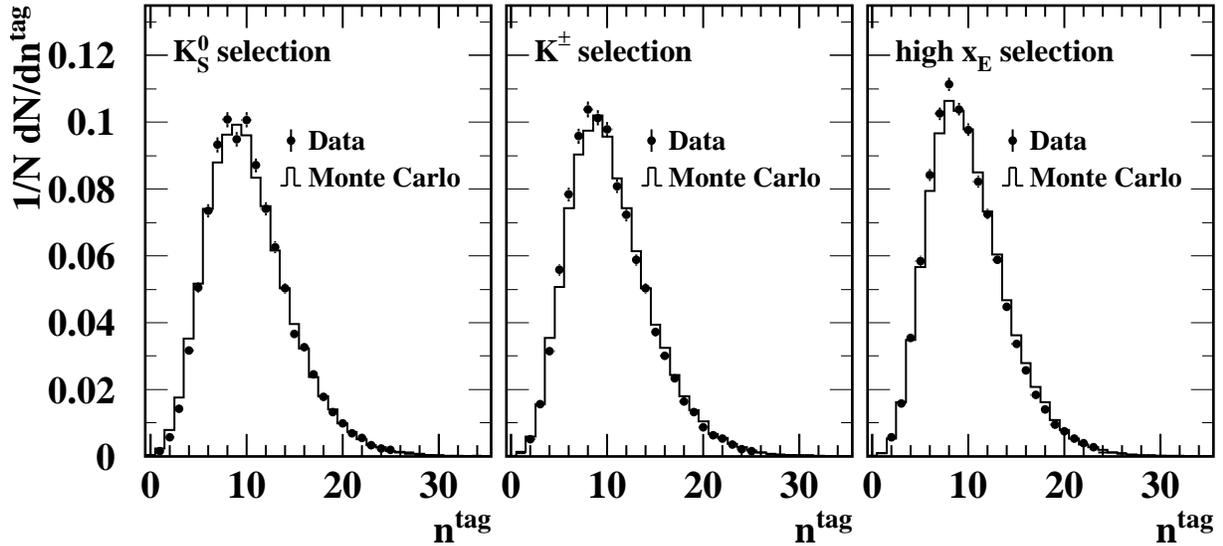}
\caption{\label{fig-tag-mult}
  Charged multiplicity distributions of the three selections, measured in the
  hemispheres opposite to the tag particles.
              Overlaid are the expectations from the 
              Monte Carlo simulation.
  }
\end{figure}
\begin{table}[b]
\begin{center}
\begin{tabular}{|l|l|r|r|r|}
\hline
         & tag         & 
                         \multicolumn{1}{c|}{\Ks} & \multicolumn{1}{c|}{\Kpm} & \multicolumn{1}{c|}{\hx} \\
\hline
\hline
Data     & 
              $\langle n^{\mathrm{tag}}\rangle$\hfill =                   
                                                           & $10.373\pm 0.031$ & $10.229\pm 0.031$ & $ 9.883\pm 0.024$ \\
MC udscb & 
              $\langle n^{\mathrm{tag}}_{\mathrm{MC}}\rangle$\hfill =     
                                                           & $10.323\pm 0.022$ & $10.362\pm 0.021$ & $10.031\pm 0.015$ \\
\hline
\hline
MC u     & 
              $\langle n^{\mathrm{tag}}_{\mathrm{u,\,MC}}\rangle$\hfill = 
                                                           & $10.171\pm 0.074$ & $10.377\pm 0.045$ & $10.089\pm 0.027$ \\
MC d     & 
              $\langle n^{\mathrm{tag}}_{\mathrm{d,\,MC}}\rangle$\hfill = 
                                                           & $10.187\pm 0.056$ & $10.370\pm 0.059$ & $ 9.995\pm 0.029$ \\
MC s     & 
              $\langle n^{\mathrm{tag}}_{\mathrm{s,\,MC}}\rangle$\hfill = 
                                                           & $10.119\pm 0.029$ & $10.242\pm 0.029$ & $ 9.921\pm 0.025$ \\
MC c     & 
              $\langle n^{\mathrm{tag}}_{\mathrm{c,\,MC}}\rangle$\hfill = 
                                                           & $10.853\pm 0.057$ & $10.684\pm 0.070$ & $10.314\pm 0.084$ \\
MC b     & 
              $\langle n^{\mathrm{tag}}_{\mathrm{b,\,MC}}\rangle$\hfill = 
                                                           & $11.932\pm 0.101$ & $11.717\pm 0.147$ & $11.660\pm 0.132$ \\
\hline
\end{tabular}
\end{center}
\caption{\label{tab-tag-mult}
  Mean charged multiplicities and statistical errors of the three selections,
  measured in the hemispheres opposite to the 
              tagged
  particles
              after applying the reweighting procedure described in
              Section~\ref{sec-check-flavour-composition}.
}
\end{table}
Figure~\ref{fig-tag-mult} shows the multiplicity distributions and
Table~\ref{tab-tag-mult} lists the averages of the distributions for the
hemisphere opposite to the tag. In the \Ks\ sample good agreement between data
and 
              the 
Monte Carlo is found, but in the 
              events selected by a highly energetic \Kpm\ or \hx\ tag
              the simulation
predicts a slightly larger multiplicity.
\subsection{Unfolding procedure}
\label{sec-unfolding}
The three tag samples are composed of events with all five primary quark
flavours
              accessible at LEP I.
A statistical unfolding procedure was applied to obtain the 
              mean
charged 
              particle
multiplicities for each light quark flavour. The procedure 
simultaneously corrects for 
              effects due to
the selection cuts and for detector effects.

The measured mean charged multiplicities,
              $\langle n^{\mathrm{tag}}\rangle$, in the hemisphere opposite to 
              the tagged particle 
are expressed as
\begin{equation}
\label{eqn-unfolding}
              \langle n^{\mathrm{tag}}\rangle=\sum_{q={\mathrm{u,d,s,c,b}}}
                 f_{q}^{\mathrm{tag}}\cdot
                 \left[\frac{\langle n_{q,{\mathrm{\,MC}}}^{\mathrm{tag}}\rangle}
                            {\langle n_{q,{\mathrm{\,MC\,had.}}}\rangle}\right]
                 \cdot\nqb\quad,\quad
                 \mathrm{tag}=\Ks,\,\Kpm,\,\hx\quad.
\end{equation}
Here $f_{q}^{\mathrm{tag}}$ is the fraction of events with primary quark flavour
$q$ in a sample denoted by ``tag''. 
              It is listed in Table~\ref{tab-fractions}.
The term in square brackets was also taken from the Monte Carlo simulation. 
It is the ratio of the mean charged multiplicity 
              of the opposite hemisphere 
for events with primary quark flavour
$q$ in the ``tag'' sample, 
              $\langle n_{q,{\mathrm{\,MC}}}^{\mathrm{tag}}\rangle$, 
              listed in Table~\ref{tab-tag-mult},
and the
number 
              for both hemispheres
obtained without detector simulation,
              $\langle n_{q,{\mathrm{\,MC\,had.}}}\rangle$, 
which is 
              given
in Table~\ref{tab-mc-gen-mult}.
\begin{table}[t]
\begin{center}
\begin{tabular}{|c|c|}
\hline
              flavour $q$ & mean multiplicity $\langle n_{q,{\mathrm{\,MC\,had.}}}\rangle$ \\
\hline
      u & $20.097\pm 0.005$ \\
      d & $19.987\pm 0.005$ \\
      s & $19.897\pm 0.005$ \\
      c & $21.387\pm 0.005$ \\
      b & $23.725\pm 0.005$ \\
\hline
\end{tabular}
\caption{\label{tab-mc-gen-mult}
  Mean 
              generated
  charged multiplicities 
              $\langle n_{q,{\mathrm{\,MC\,had.}}}\rangle$
  as obtained from the Monte Carlo generator JETSET
  7.4. Only charged particles with life-times $\tau>300$\,ps were considered.}
\end{center}
\end{table}
This ratio corrects (i) for the biases due to the tagging cuts, (ii) for the
restriction to the hemisphere opposite to the tag particle when determining the
charged multiplicity, and (iii) for effects from the selection of multihadronic
events, the finite resolution and acceptance of the detector. On average the
correction factor in square brackets is $1/(2\times 1.007)$ for all flavours varying
              at most by $\pm 3.7\,\%$, and 
where the factor of 2 is due to considering a single
hemisphere only for the determination of the multiplicities denoted by ``tag''.

The flavour dependent mean charged multiplicities, \nqb, can be determined by
solving the equation system~(\ref{eqn-unfolding}). The values of \ncb\ and \nbb\
were taken from our previous determination~\cite{bib-OPAL-EPJC7-369}:
\begin{eqnarray}
\ncb & = & 21.55\pm 0.37\pm 0.64 \\
\nbb & = & 23.16\pm 0.02\pm 0.45 \quad,
\end{eqnarray}
where the errors are statistical and systematic, respectively. The solution of
the equation system~(\ref{eqn-unfolding}) using the measured multiplicities in
the hemisphere opposite to the 
              tagging particles from Table~\ref{tab-tag-mult} yielded
              
\begin{eqnarray*}
\nub & = & 17.77\pm 0.51 \\
\ndb & = & 21.44\pm 0.63 \\
\nsb & = & 20.02\pm 0.13
\end{eqnarray*}
where the errors are statistical only. 
             Combining these, and considering the statistical correlation matrix:
\begin{center}
\begin{tabular}{c|rrr}
  & \multicolumn{1}{c}{u} & \multicolumn{1}{c}{d} & \multicolumn{1}{c}{s} \\
\hline
u & $1.00$ & $-0.90$ & $+0.06$ \\
d &        &  $1.00$ & $-0.42$ \\
s &        &         &  $1.00$ \\
\end{tabular}
\end{center}
             and using the branching ratios of $\PZ\rightarrow 
             \mathrm{u\overline{u}}$, $\mathrm{d\overline{d}}$, 
             $\mathrm{s\overline{s}}$ 
              from~\cite{bib-ZFITTER} one obtains
              $\langle n_{\mathrm{uds}}\rangle=19.90 \pm 0.09\ {\mathrm{(stat.)}}$.
             This is in agreement within systematic errors with the 
              results 
              of~\cite{bib-OPAL-EPJC7-369,bib-OPAL-EPJC1-479} 
             which were obtained using a different method.
              The mean charged multiplicities 
\nub\ and \ndb\ are almost fully anti-correlated, since the 
              statistical separation power for u and d quarks is due to 
              the small but different fractions of these quarks 
              in the \Ks\ and the \Kpm\ samples.

              \subsection{Test of the unfolding procedure}
              \label{sec-check-unfolding}
              It was checked that the unfolding procedure described in 
              Section~\ref{sec-unfolding} is able to determine the
              flavour dependent multiplicities. For this test the 
              Monte Carlo data sample was split into two disjoint parts. 
              The smaller, comprising about the same statistics as 
              the measured data, was used in place of the measured data 
              which were to be corrected using the larger part of the 
              Monte Carlo events. The multiplicities 
              $\nub=20.05\pm0.63\,(\mathrm{stat.})$,
              $\ndb=19.98\pm0.81\,(\mathrm{stat.})$,
              and
            $\nsb=19.89\pm0.16\,(\mathrm{stat.})$
            are in excellent agreement with the mean multiplicities 
            obtained from the Monte Carlo generator
            (cp.\ Table~\ref{tab-mc-gen-mult}).
%
%
%
              \section{Systematic study of the particle multiplicities}
\label{sec-systematics}
In addition to the consistency check based on the double tag rates
in Section~\ref{sec-check-flavour-composition} and the
              test of the unfolding procedure in 
              Section~\ref{sec-check-unfolding}, many sources of systematic
              uncertainties were investigated which
may be subdivided into five groups, namely (i) fluctuations due to limited 
Monte Carlo statistics, (ii) the experimental precision of the multiplicity 
in charm and bottom events, (iii) impacts of detector 
              effects and event selection and 
              the   dependence
              of the mean charged multiplicity
on the choice of the hadronization model, (iv) variations of 
tagging cuts,
and finally, (v) 
              impacts on the flavour fractions,
              $f_q^{\mathrm{tag}}$,
           due to variations of 
              hadronization parameters of the Monte Carlo generator, 
              the matching of the Monte Carlo $x_E$ spectra and 
              hemisphere correlations to the data, and changes of the 
              hadronization model.
Apart from 
the items (i) and (ii), whose error contributions were estimated 
by error propagation, the general procedure to derive the uncertainty associated 
with the source was 
              to repeat
the complete analysis with one of the cuts or parameters varied. Any deviation 
from the result found for the standard set of cuts and parameters was interpreted 
as a systematic uncertainty.

The respective error contributions of the uncertainties derived for the groups
              (i)
to (v) were quadratically added to estimate the total systematic
uncertainties. The individual error contributions are listed in
Table~\ref{tab-syst-errors}.
\begin{table}
\begin{center}
\renewcommand{\arraystretch}{1.25}
\begin{tabular}{|l|l||l|l|l|}
\hline
             \multicolumn{2}{|l||}{source of uncertainty}&
\multicolumn{1}{c|}{$\Delta\nub$} & \multicolumn{1}{c|}{$\Delta\ndb$} & \multicolumn{1}{c|}{$\Delta\nsb$} \\
\hline\hline
             \multicolumn{2}{|l||}{data statistics}      & $\mathbf{\pm 0.51}$ & $\mathbf{\pm 0.63}$ & $\mathbf{\pm 0.13}$ \\
\hline\hline
             \multicolumn{2}{|l||}{MC statistics}        & $\pm 0.37$          & $\pm 0.48$          & $\pm 0.09$          \\
\hline\hline
\multicolumn{2}{|l||}{c, b multiplicity~\cite{bib-OPAL-EPJC7-369}}
                                                         & $\pm 0.27$          & $\pm 0.13$          & $\pm 0.19$          \\
\hline\hline
\multicolumn{2}{|l||}{detector~\cite{bib-OPAL-mult}}
                                                         & $\pm 0.17$          & $\pm 0.20$          & $\pm 0.19$          \\
\hline
\hline
& $x_E$-cut variation                                    & \apm{-0.16}{+0.49}  & \apm{-0.79}{+0.35}  & \apm{+0.22}{-0.10}  \\
              \raisebox{2mm}[0mm][0mm]{
              \rotatebox{90}{\hspace*{0pt\hfill}all\hspace*{0pt\hfill}}}
& $x_E<0.9$                                              & $-0.06$             & $-0.06$             & $+0.02$             \\
\hline
              \raisebox{-1mm}[0mm][0mm]{
              \rotatebox{90}{\hspace*{0pt\hfill}\Ks\hspace*{0pt\hfill}}}
& \Ks\ mass window                                       & $-0.24$             & $+0.22$             & $+0.04$             \\
              \hline
& $|\cos\theta_{\Kpm}|<0.6$, $0.8$                       & $\pm 0.14$          & $\mp 0.20$          & $\pm 0.04$          \\
& $N_{\mathrm{CJ}}>80$, $N_{\dedx}>80$                   & $+0.09$             & $-0.14$             & $+0.02$             \\
& $N_{\dedx}>110$                                        & $-0.15$             & $+0.22$             & $-0.04$             \\
              \raisebox{9mm}[0mm][0mm]{
              \rotatebox{90}{\hspace*{0pt\hfill}\Kpm\hspace*{0pt\hfill}}}
& $|w_{\Kpm}|\ge 0.5$, $w_{\mathrm{p}}>0.4$, $w_{\pi^{\pm}}>-0.06$
                                                         & $+0.16$             & $-0.23$             & $+0.04$             \\
              \hline
& $|\cos\theta_{\hx}|<0.6$, $0.8$                        & \apm{+0.01}{-0.01}  & \apm{+0.11}{-0.16}  & \apm{+0.05}{-0.03}  \\
& $N_{\mathrm{CJ}}>120$, $140$                           & $<0.01$             & \apm{+0.03}{-0.01}  & $\mp 0.01$          \\
              \raisebox{2mm}[0mm][0mm]{
              \rotatebox{90}{\hspace*{0pt\hfill}\hx\hspace*{0pt\hfill}}}
& $\chi^2_{r\varphi}<1.5$                                & $+0.01$             & $+0.11$             & $-0.03$             \\
\hline
             \multicolumn{2}{|l||}{selection cuts total} & \apm{+0.54}{-0.36}  & \apm{+0.53}{-0.88}  & \apm{+0.24}{-0.12}  \\
\hline
\hline
\multicolumn{2}{|l||}{$\Lambda_{\mathrm{LLA}}=250\pm 6$\,MeV}
                                                         & $\pm 0.04$          & $\pm 0.06$          & $\pm 0.01$          \\
\multicolumn{2}{|l||}{$Q_0=1.9\pm 0.5$\,GeV}             & $\pm 0.03$          & $\pm 0.05$          & $\pm 0.03$          \\
\multicolumn{2}{|l||}{$\sigma_q=0.40\pm 0.03$\,GeV}      & $\pm 0.04$          & $\pm 0.05$          & $\pm 0.01$          \\
\multicolumn{2}{|l||}{$b=0.52^{+0.40}_{-0.26}$}          & \apm{-0.06}{+0.12}  & \apm{+0.22}{-0.20}  & \apm{+0.04}{-0.09}  \\
\multicolumn{2}{|l||}{$\varepsilon_{\mathrm{c}}=0.031\pm 0.011$}
                                                         & $\pm 0.06$          & $\pm 0.06$          & $\pm 0.05$          \\
\multicolumn{2}{|l||}{$\varepsilon_{\mathrm{b}}=0.0038\pm 0.0010$}
                                                         & $\pm 0.06$          & $\pm 0.11$          & $\pm 0.02$          \\
\multicolumn{2}{|l||}{$\gamma_{\mathrm{s}}=0.31\pm 0.10$}& \apm{+0.30}{-0.68}  & \apm{-0.34}{+0.74}  & \apm{+0.06}{-0.07}  \\
\multicolumn{2}{|l||}{$\mathrm{(us/ud)/(s/d)}=0.45\pm 0.04$} 
                                                         & $\pm 0.03$          & $\pm 0.06$          & $\pm 0.02$          \\
\multicolumn{2}{|l||}{$V_{\mathrm{d,u}}=0.60 \pm 0.10$}  & $\pm 0.04$          & $\pm 0.07$          & $\pm 0.02$          \\
\multicolumn{2}{|l||}{$V_{\mathrm{s}}=0.40\pm 0.05$}     & \apm{-0.36}{+0.26}  & \apm{+0.39}{-0.31}  & \apm{-0.04}{+0.03}  \\
\multicolumn{2}{|l||}{no tensor mesons}                  & $-0.58$             & $+0.85$             & $-0.14$             \\
\multicolumn{2}{|l||}{$B({\mathrm{c}}\rightarrow \Ks  + X): \pm 13.9\,\% \oplus B({\mathrm{b}}\rightarrow \Ks  + X): \pm 10.0\,\%$}
                                                         & $\pm 0.14$          & $\mp 0.13$          & $\mp 0.02$          \\
\multicolumn{2}{|l||}{$B({\mathrm{c}}\rightarrow \Kpm + X): \pm 11.5\,\% \oplus B({\mathrm{b}}\rightarrow \Kpm + X): \pm 21.6\,\%$}
                                                         & $\mp 0.07$          & $\pm 0.10$          & $\pm 0.02$          \\
\multicolumn{2}{|l||}{reweighted $x_E$ spectra}          & $-0.35$             & $+0.17$             & $-0.03$             \\
\multicolumn{2}{|l||}{hemisphere correlations}           & $-0.01$             & $+0.02$             & $<0.01$             \\
\hline
\multicolumn{2}{|l||}{hadronization total}               & \apm{+0.46}{-1.05}  & \apm{+1.25}{-0.56}  & \apm{+0.11}{-0.20}  \\
\hline
\hline
\multicolumn{2}{|l||}{total systematic errors} &
                                                          \apm{\mathbf{+0.86}}{\mathbf{-1.20}} 
                                                                      & \apm{\mathbf{+1.46}}{\mathbf{-1.17}} 
                                                                                    & \apm{\mathbf{+0.39}}{\mathbf{-0.37}} \\
\hline
\hline
\multicolumn{2}{|l||}{total errors} &
                                                          \apm{\mathbf{+1.01}}{\mathbf{-1.31}}
                                                                      & \apm{\mathbf{+1.59}}{\mathbf{-1.33}}
                                                                                    & \apm{\mathbf{+0.41}}{\mathbf{-0.40}} \\
\hline
\end{tabular}
\end{center}
\caption{\label{tab-syst-errors}
  Compilation of individual contributions to the total error.}
\end{table}
In the following sections the sources (iii) to (v) of systematic uncertainties
will be discussed in more detail.
\subsection{Uncertainties from detector simulation and event selection}
\label{sec-detector-syst}
The detector simulation was needed to correct for the fraction of multiplicity
that was not recorded by the detector either owing to limited acceptance, biases
of the event selection, or interactions in the detector 
              such as
$\delta$-electrons or hadronic interactions, all of which have been studied in
great detail 
              in~\cite{bib-OPAL-mult}. In 
total an error of about
1\,\% due to this source of systematic uncertainty was 
              estimated in~\cite{bib-OPAL-mult}. 
This was also adopted for this measurement.
\subsection{Uncertainties from tagging}
\label{sec-tagging-syst}
The measured multiplicity depends on the energy of the tagged particle. Although
this effect is reduced by measuring the multiplicity in the hemisphere opposite
to the tag residual effects have to be taken into account. 
              \begin{itemize}
              \item
Variations of the
$x_E$ cuts in steps of 0.05 by as much as $\pm 0.1$ around the default cut value
were done. The largest up- and downward excursions yielded the error estimates for
the multiplicities.
              \item
Also a more stringent cut on the maximum accepted 
              momentum, $x_E < 0.9$,
for all tags did not change the measured multiplicities by more than 0.4\,\%.
              \end{itemize}
For each of the three tag types the impact of the choice of the most important
selection cuts was studied. 
              \begin{itemize}
              \item
For the \Ks\ the allowed mass window around the
world average mass value~\cite{bib-PDG} was altered from 120\,MeV 
within the
range of 80 to 200\,MeV.
              \end{itemize}
Since the \Kpm\ selection relies very much on the capability to precisely
measure the energy loss in the jet chamber, the relevant cuts were investigated.
              \begin{itemize}
              \item
As the precision of the measurement of \dedx\ degrades towards the very forward
and backward regions, the $|\cos\theta_{\Kpm}|<0.72$ requirement was varied in
small steps between 0.6 and 0.8, taking the r.m.s.\ to estimate the contribution
to the systematic uncertainty. 
              \item
The minimum number of both jet chamber hits and 
\dedx\ hits 
              was relaxed from 130
to 80.
              \item
              The 
required number of \dedx\ hits 
              was increased from 100
to 110.  
              \item
From the various modifications to the energy loss weight cuts, namely 
choosing a tighter cut on the kaon \dedx\ weight of 0.5 instead of 0.1, adding 
an additional rejection cut on the proton \dedx\ weight of 0.4, or considering 
a better pion rejection by requiring its \dedx\ weight to be larger than $-0.06$, 
only the first one contributes a significant systematic error.
              \end{itemize}
The \hx\ selection critically depends on the ability of the jet chamber to
reliably measure high momentum particles. 
              \begin{itemize}
              \item
As for \Kpm, the allowed range of 
$|\cos\theta_{\hx}|$ was varied from less than 0.72 to either less than 0.6 
or 0.8.  
              \item
A variation of the minimum number of hits required for a \hx\ particle 
in the jet chamber by $\pm 10$ around the standard value of 130 had only a 
negligible effect on all flavours. 
              \item
Tightening the cut on the $\chi^2_{r\varphi}$ 
of the track fit from 2 to 1.5 affected the multiplicity by less than 0.5\,\%.
              \end{itemize}
              \subsection{Uncertainties due to impacts on flavour fractions}
\label{sec-hadronization-syst}
The Monte Carlo event generator requires several parameters to be adjusted for a
proper description of the measured data. Changing these hadronization parameters
             affects
the flavour composition,
              $f_q^{\mathrm{tag}}$,
and, therefore, tests the 
              sensitivity of the result of this analysis
on the particular choice of these parameters. In the following we
consider those parameters which are expected to have an impact on the flavour
fractions. These were varied about their tuned 
              values
within the intervals quoted 
              in~\cite{bib-OPALtune}  
              apart from 
               from the $b$ and $\gamma_s$ parameters
               where larger variations
               were investigated,
               in particular the range for $\gamma_s$ given 
               in~\cite{bib-CERN-9601}. 
Most of the parameter variations did not
significantly affect the flavour dependent 
              multiplicities, $\langle n_{q,{\mathrm{\,MC\,had.}}}\rangle$,
              or
the flavour 
              fractions, $f_{q}^{\mathrm{tag}}$, of Eq.~(\ref{eqn-unfolding}).
We therefore mention only the significant changes of 
the flavour 
              fractions
due to the parameter variations. 
              \begin{itemize}
\item The amount of gluon radiation was modified by varying the 
      $\Lambda_{\mathrm{LLA}}$ parameter and the cut-off of the 
      parton shower, $Q_0$. 
\item During the hadronization step a hadron receives extra
      transverse momentum whose size is controlled by the 
      parameter $\sigma_q$ which was changed from its tuned value. 
\item The fraction of energy and momentum transferred from a heavy 
      c or b quark to the related hadron also has an impact on the 
      background contributions from these heavy quarks. The relevant
      $\varepsilon$ parameters of the Peterson et al.~\cite{bib-Peterson-fragmentation}
      fragmentation function were varied. Only the change of
      $\varepsilon_{\mathrm{c}}$, except for the charm quark 
      fraction itself, had a noticeable effect on the strange quark 
      fraction of about $-0.01$ to $+0.005$ for the two kaon selections.
\item To account for a remaining uncertainty from the hardness of the light quark
      fragmentation functions despite the impact of the reweighting procedure, the
      parameter $b$ of the Lund symmetric fragmentation function in JETSET was varied
      in several steps in the range of 0.26 to 0.92. This changed in particular the
      strange quark fraction in the range of $-0.06$ to $+0.04$ for the \Ks\ and in
      the range of $-0.04$ to $+0.03$ for the \Kpm\ selection. The u and d fractions
      varied owing to this variation of $b$ by $\pm 0.012$ at most in all three
      selections.
\end{itemize}

             \begin{itemize}
\item Since the statistical separation between u and d quarks is based on the \Ks\
      and \Kpm\ production, the relative amount of \ssbar\ quark pairs created from
      the vacuum, $\gamma_s$, was changed in steps in the range of 0.21 to
      0.41~\cite{bib-CERN-9601}. 
                   This led to changes of the light quark fractions in the \Kpm\ (\Ks)
                   selection ranging from 
                   $-0.05$, $-0.03$, $+0.06$ ($-0.02$, $-0.03$, $+0.04$)
                   to
                   $+0.06$, $+0.02$, $-0.05$ ($+0.01$, $+0.02$, $-0.04$),
                   respectively, for u, d, s
              quarks.
              For the \hx\ selection only changes of the u and s quark fractions were
              observed at a level of $\pm 0.01$.
\item For baryons with strangeness an extra suppression factor $\mathrm{(us/ud)/(s/d)}$ 
      exists in JETSET whose variation resulted in only negligible changes of both the 
      flavour composition and the flavour dependent multiplicities. 
\item The 
              production
      of vector 
      mesons was considered. Although the down 
      and strange quark fractions of the \hx\ selection changed by about $\pm 0.02$, 
      the relative portion $V_{\mathrm{u,d}}$ of vector mesons from u and d quarks 
      was of minor importance for the flavour dependent multiplicities. 
\item The fraction of vector mesons from s quarks, $V_{\mathrm{s}}$, substantially 
              affects
      the yields of 
               leading
      \Ks\ and \Kpm\ mesons. The variation of this parameter
      mostly changed the fraction from up and down quarks 
              in
      the \Kpm\ selection by
      about $\pm 0.01$, and at the same level also that from down and strange quarks
      of the \hx\ selection. Since the 
              statistical u-d quark
      separation strongly relies on the u and d
              quark
      fractions in the two kaon selections, a change of this parameter led to a large 
      error contribution as can be seen from Table~\ref{tab-syst-errors}. 
\item Similarly, the decay products of tensor mesons have a softer $x_E$ spectrum. 
      However, the yields of these are not well known. Thus, the impact was estimated 
      by analyzing about 3.5 million fully simulated events from the JETSET generator 
      tuned to describe the data, but without the production of tensor mesons. This 
      yielded a substantial reduction of the strange quark fraction by $-0.025$ in 
      the \Ks, $-0.035$ in the \Kpm, and $-0.065$ in the \hx\ 
                    selections.
      The up and down 
              quark
       fractions in the \Ks\ and \Kpm\ selections changed by $-0.01$ 
      and $-0.005$, whereas the d 
              quark
      fraction in the \hx\ selection increased by about 
      $+0.03$. Thus the d and s 
              quark
      fractions in the \hx\ selection are about equal if 
      tensor mesons do not contribute. The difference in the results with simulated 
      tensor mesons is assigned as a systematic error.
\item The
      uncertainty of the inclusive branching ratios of charm and bottom to \Ks\ 
      and \Kpm\ was considered. From the data 
              in~\cite{bib-PDG} 
      the
      relative errors on the $B({\mathrm{c}} \rightarrow \Ks X, \Kpm X)$ branching
      ratios have been determined to be $\pm 11.5\,\%$, $\pm 13.9\,\%$, and on
      $B({\mathrm{b}} \rightarrow \Ks X, \Kpm X)$ to be $\pm 21.6\,\%$,
      $\pm 10.0\,\%$, respectively. Their impact on the flavour dependent
      multiplicities was assessed by varying the charm and bottom contributions 
      in the \Ks\ and \Kpm\ tagged 
              event
      samples according to these percentages.
\item To account for the poor simulation of the $x_E$ spectra of the 
      tagged particles in the 
              tuned
      Monte Carlo the reweighting functions 
      of Section~\ref{sec-check-flavour-composition}, 
      $f(x_E)=c\cdot(1-x_E)^{\delta}$, were employed. Applying these specific 
      reweighting functions to all three selections resulted in a small 
      reduction of the strange quark fraction by 1 to 2\,\% while the 
      heavy quark contribution is increased accordingly. Moreover,
      it leads to an anti-correlated change of the order of 1 to 3\,\%
      of the u and d 
              quark
      fractions but in opposite directions for the \Ks\
      and the \Kpm\ 
              event samples.
      The final multiplicities of u
      and d quarks change by up to 2\,\%.
\item Finally, the small discrepancies between simulation and data noticed in the
      correlations of the opposite hemispheres shown in Figure~\ref{fig-hem-corr} 
      were considered. Depending on the $x_E$ of the 
              tagged particle
      the Monte Carlo events 
      were reweighted to match the data. This had only a negligible effect on 
      both the flavour fractions and the flavour dependent multiplicities 
      changing them by less than 0.1\,\%.
              \end{itemize}
              \par
In addition to the JETSET Monte Carlo event generator also 
      the HERWIG program~\cite{bib-HERWIG} was examined. While HERWIG yields a 
              good
      description of the shape of the $x_E$ spectra of the tagged 
      particles, it 
              does not reproduce 
      the absolute rates. For instance, 
              HERWIG
      predicts a much larger contribution 
             to the \hx\ sample 
      from d quarks and less from u quarks 
      due to 
              a larger number of protons produced
      in HERWIG.
      Since the HERWIG expectation of the total proton yield is not in agreement with
      the measurements~\cite{bib-Green-book,bib-PDG}, one might suspect that the
      flavour fractions obtained from this Monte Carlo generator are not reliable.
      Thus, no systematic error contribution from using HERWIG is quoted.
%
%
%
%
             \section{Cross-check with flavour fractions derived from double tag counts}
\label{sec-fitted-fractions}
              The
flavour 
              fractions, $f_{q}^{\mathrm{tag}}$, can be derived from data only using
the 
              number of
single and double tag counts since they
are related to the sum 
              and products of flavour fractions 
              \cite{bib-Letts-Maettig}. 
The non-linearity of the equation system requires for its solution more than the 
six relations which are given by the numbers of single and double tags considered 
in this analysis even when the heavy flavour contributions are kept fixed.

Such an investigation has been done 
              in~\cite{bib-OPAL-ZPC76-387}
              using
the single and double tag counts of charged pions, kaons,
protons, \Ks\ and $\Lambda$ for $x_E>0.4$, $0.5$, and $0.6$.
              This analysis yielded
the flavour dependent production 
rates of these hadrons where only the SU(2) isospin flavour symmetry and the 
flavour independence of 
              the strong interaction
were assumed for the production of charged pions, 
and charged and neutral kaons. 

              To
compare with the results from 
              \cite{bib-OPAL-ZPC76-387}, 
the flavour 
              fractions, $f_{q}^{\mathrm{tag}}$,
were calculated from the quoted production rates
                           neglecting correlations 
              for this cross-check. 
The calculation took into account that these rates had been corrected for 
mis-identification of the tagging particle. The \hx\ sample was approximated 
by adding the production rates of the charged particles, i.e.\ pions, kaons, 
and protons. The comparison of the flavour fractions at an $x_E$-cut of 
              0.4 for \Ks, 0.5 for \Kpm\ and 0.6 in the case of the \hx\ sample
revealed some substantial differences with respect to the JETSET prediction. The
\Ks\ fraction from up quarks is $0.4$ times and that from down quarks $1.7$ times
the value obtained from the Monte Carlo simulation used for this analysis. The 
up quark contribution to the \Kpm\ sample is 
             1.23 times
higher, while 
              that of down quarks is
              about 
              0.75 times that
obtained from the simulation.
The 
fraction of \hx\ particles originating from up quarks is about 10\,\% less than
expected from the simulation. The strange quark fractions agree within $\pm 0.02$ 
for each 
               tagged event
sample.  These changes in the 
              up and down quark
flavour fractions are considerably larger than what was found from the variation 
of the JETSET parameters in Section~\ref{sec-hadronization-syst}.

Although the analysis 
              in~\cite{bib-OPAL-ZPC76-387} 
explicitly assumed
flavour independence of the strong coupling and SU(2) isospin symmetry, the flavour 
fractions obtained 
              from~\cite{bib-OPAL-ZPC76-387} 
were used for the unfolding procedure
described in Section~\ref{sec-unfolding}
              as a cross-check of the results.
              This 
yielded changes of the flavour dependent multiplicities of $\Delta\nub=+0.60$,
$\Delta\ndb=-1.41$, and $\Delta\nsb=+0.32$. 
              Since these changes are essentially covered by the systematic
              errors discussed in Section~\ref{sec-systematics} they are
              not considered as an additional systematic uncertainty.

%
%
%
%
%
\section{Results}
\label{sec-result-syst}
Adding the individual contributions to the systematic uncertainty from 
Monte Carlo statistics, 
              the mean charged multiplicity in c and b quark events, the 
correction of detector acceptance and resolution, 
              the
selection cuts, and 
              the
parameters of the hadronization 
model in quadrature, 
              the total systematic error 
              of the flavour dependent mean charged multiplicities is
              about $\pm 7$\,\% for up and down, and $\pm 2$\,\% for strange quark events.
The final results are
\begin{eqnarray*}
              \nub & = & \mathrm{17.77\pm 0.51\,(stat.)\,
                   ^{\displaystyle+0.86}_{\displaystyle-1.20}\,(syst.)}\\[1ex]
              \ndb & = & \mathrm{21.44\pm 0.63\,(stat.)\,
                                 ^{\displaystyle+1.46}_{\displaystyle-1.17}\,(syst.)}\\[1ex]
              \nsb & = & \mathrm{20.02\pm 0.13\,(stat.)\,
                   ^{\displaystyle+0.39}_{\displaystyle-0.37}\,(syst.)}\,.
\end{eqnarray*}
All errors apart from selection 
              cuts, detector effects, and hadronization models
were obtained 
by error propagation and, therefore, obey the statistical correlation as 
quoted in Section~\ref{sec-unfolding}. The correlation coefficients for the 
remaining two groups of systematic uncertainties were determined 
              separately for each individual systematic variation.
              \par
              These flavour dependent mean charged multiplicities agree within
              the total errors, even though deviations may be expected due to
              decays of hadrons. For instance, the decay of the $\phi(1020)$ to 
              charged kaons is enhanced over the decay to neutral kaons.
            Thus, the results from the 
              JETSET generator displayed in Table~\ref{tab-mc-gen-mult} show 
              a $1\%$ spread between the charged multiplicities from up and 
              strange quark events.

              Some of the flavour dependent mean charged multiplicities are strongly
              correlated which must be considered when comparing the results. 
              Taking these statistical and systematic correlations into 
              account the ratios of the multiplicities are 
\begin{eqnarray*}
\frac{\nub}{\ndb} & = &
              0.829 \pm 0.047\,{\mathrm{(stat.)}}
      ^{\displaystyle{+0.081}}_{\displaystyle{-0.109}}\,{\mathrm{(syst.)}}\,,\\
\frac{\nsb}{\ndb} & = &
              0.934 \pm 0.030\,{\mathrm{(stat.)}}
      ^{\displaystyle{+0.062}}_{\displaystyle{-0.089}}\,{\mathrm{(syst.)}}\,,\\
\frac{\nsb}{\nub} & = &
1.127 \pm 0.033\,{\mathrm{(stat.)}}
      ^{\displaystyle{+0.057}}_{\displaystyle{-0.077}}\,{\mathrm{(syst.)}}\,,
\end{eqnarray*}
              which are all close to unity.
%
%
%
\section{Summary and conclusions}
\label{sec-conclusions}
A measurement of the charged multiplicity in events of u, d, and s quarks from
\epem\ annihilation at the Z pole is presented. Exploiting the leading particle
effect to identify the primary quark flavour, samples of events differently
enriched in u, d, and s were selected by tagging highly energetic \Ks, \Kpm\ and
charged particles. The multiplicities per light quark flavour were obtained from
a statistical unfolding of the multiplicities measured in the hemispheres
opposite to the tagged particles. Hemisphere correlations were observed, but
were found to be 
              small
and well-described by the Monte Carlo simulation. The
final corrected multiplicities were determined to be
\begin{eqnarray*}
             \nub & = & \mathrm{17.77 \pm 0.51\,(stat.)\,
                                ^{\displaystyle +0.86}_{\displaystyle -1.20}\,(syst.)}\\[1ex]
             \ndb & = & \mathrm{21.44 \pm 0.63\,(stat.)\,
                                ^{\displaystyle +1.46}_{\displaystyle -1.17}\,(syst.)}\\[1ex]
             \nsb & = & \mathrm{20.02 \pm 0.13\,(stat.)\,
                                ^{\displaystyle +0.39}_{\displaystyle -0.37}\,(syst.)},
\end{eqnarray*}
where the 
             $\nub$ and $\ndb$
multiplicities are highly statistically anti-correlated (about
              $-90\,\%$). 
              The ratios of pairs of these multiplicities which take the correlations
              into account are all close to unity. This agrees with the expectation
              of QCD that the charged particle multiplicities are flavour independent
              for the light up, down, and strange quarks
              apart from small effects due to particle decays.


\appendix
\par
Acknowledgements:
\par
We particularly wish to thank the SL Division for the efficient operation
of the LEP accelerator at all energies
 and for their continuing close cooperation with
our experimental group.  We thank our colleagues from CEA, DAPNIA/SPP,
CE-Saclay for their efforts over the years on the time-of-flight and trigger
systems which we continue to use.  In addition to the support staff at our own
institutions we are pleased to acknowledge the  \\
Department of Energy, USA, \\
National Science Foundation, USA, \\
Particle Physics and Astronomy Research Council, UK, \\
Natural Sciences and Engineering Research Council, Canada, \\
Israel Science Foundation, administered by the Israel
Academy of Science and Humanities, \\
Minerva Gesellschaft, \\
Benoziyo Center for High Energy Physics,\\
Japanese Ministry of Education, Science and Culture (the
Monbusho) and a grant under the Monbusho International
Science Research Program,\\
Japanese Society for the Promotion of Science (JSPS),\\
German Israeli Bi-national Science Foundation (GIF), \\
Bundesministerium f\"ur Bildung und Forschung, Germany, \\
National Research Council of Canada, \\
Research Corporation, USA,\\
Hungarian Foundation for Scientific Research, OTKA T-029328, 
T023793 and OTKA F-023259.\\


%
\end{document}